# Gait complexity assessed by detrended fluctuation analysis is sensitive to inconsistencies in stride time series: A modeling study

Philippe Terrier[a,b,§]

[a] Haute-Ecole Arc Santé, HES-SO University of Applied Sciences and Arts Western Switzerland, Neuchâtel, Switzerland

[b] Department of Thoracic and Endocrine Surgery, University Hospitals of Geneva, Geneva, Switzerland

[§]**Corresponding author:**

Philippe Terrier

Haute Ecole Arc Santé

Espace de l'Europe 11, CH-2000 Neuchâtel, Switzerland

Email address: Philippe.terrier@he-arc.ch

**Abstract**

*Background:* Human gait exhibits complex fractal fluctuations among consecutive strides. The time series of gait parameters are long-range correlated (statistical persistence). In contrast, when gait is synchronized with external rhythmic cues, the fluctuation regime is modified to stochastic oscillations around the target frequency (statistical anti-persistence). To highlight these two fluctuation modes, the prevalent methodology is the detrended fluctuation analysis (DFA). The DFA outcome is the scaling exponent, which lies between 0.5 and 1 if the time series exhibit long-range correlations, and below 0.5 if the time series is anti-correlated. A fundamental assumption for applying DFA is that the analyzed time series results from a time-invariant generating process. However, a gait time series may be constituted by an ensemble of sub-segments with distinct fluctuation regimes (e.g., correlated and anti-correlated).

*Research question:* To what extent do non-constant time series modify the scaling exponent computed via DFA?

*Methods:* Several proportions of correlated and anti-correlated time series were mixed together and then analyzed through DFA. The original (before mixing) time series were generated via autoregressive fractionally integrated moving average (ARFIMA) modelling or actual gait data.

*Results:* Results evidenced a nonlinear sensitivity of DFA to the mix of correlated and anti-correlated series. Notably, adding a small proportion of correlated segments into an anti-correlated time series had stronger effects than the reverse. For example, integrating 20% of correlated segments (scaling exponent: 0.8) within an anti-correlated series (scaling exponent: 0.3) resulted in a time series interpreted as random (scaling exponent 0.5).

*Significance:* In case of changes in gait control during a walking trial, the resulting time series may be a patchy ensemble of several fluctuation regimes. When applying DFA, the scaling exponent may be misinterpreted. Cued walking studies may be most at risk of suffering this issue in cases of sporadic synchronization with external cues.

# 1. Introduction

Human walking is a complex motor task involving both proactive and reactive adaptations. The proactive processes rely on internal neural models and pattern generators (feedforward control), and the reactive processes rely on sensorimotor and visual inputs (feedback control). This intricate neural organization allows for an optimal response to environmental demands [1]. As a result, gait variability exhibits fractal features, which is one of the hallmarks of complex physiological signals [2]. It is thought that such a characteristic points to the ability to flexibly adapt [3]. Therefore, the fluctuations of gait parameters while walking are not insignificant; on the contrary, the fluctuation structure conveys important information regarding the system behavior. The consequence of this scale-free, fractal organization of the gait is that stride lengths, durations, and speeds vary from cycle to cycle with long-range correlations (statistical persistence) among dozens of consecutive strides [4,5]; that is, a short stride is more likely to be followed by another short stride, and a long stride by another long stride.

Sensorimotor synchronization refers to the capability of coordinating movements with external rhythms [6]. Recent insights into the neural basis of motor control suggest that anticipatory rhythmic templates in the brain allow for motor planning and synchronized movement execution with an external rhythmic cue [7]. In motor rehabilitation, rhythmic auditory cueing can be exploited to reprogram the execution of a motor pattern and reinstates normal movements [7]. Rhythmic auditory cueing has proven to be efficient in treating movement disorders in Parkinson's disease [8] and stroke [9].

Regarding gait variability, rhythmic auditory cueing induces a specific pattern of stride-to-stride fluctuations. When the gait is paced, the long-range correlated—fractal— time series change to anti-correlated ones (statistical anti-persistence). That is, a short stride is more likely to be followed by a long stride. In long durations of walking paced by a metronome, stride durations stochastically oscillate around the target frequency, while stride lengths and speeds remain fractal [5]. A similar response is observed for stride speed series in treadmill experiments; here, the reference is the treadmill speed; the actual speed of the walking individual oscillate from stride to stride around the reference speed with an anti-correlated pattern [10]. When performing auditory cueing experiments on a treadmill, stride speeds, lengths, and durations become anti-persistent [10]. Similar effects can be induced when step lengths are aligned with visual cues projected onto the treadmill [11]. In contrast to normal (not paced) walking—in which small deviations can persist across strides—cued walking requires a tight control of the gait with rapid corrections of deviations away from the target. The corrections are often slightly too large, which induces, to compensate, corrections in the opposite direction, resulting in the observed statistical anti-persistence [11,12].

The most popular algorithm for characterizing the fluctuation structure of gait variability is detrended fluctuation analysis (DFA) [13]. DFA reports the log of variance in time series as a function of the log of time scales (power law): The slope of the linear fit between these parameters is the scaling exponent. The scaling exponent lies between 0.5 and 1 if the time series exhibits long-range correlations, and below 0.5 if the time series is anti-correlated. A scaling exponent of 0.5 indicates a random—uncorrelated—time series. When DFA is applied for characterizing gait fluctuations, a substantial number of consecutive strides must be included in the analysis [14]. This is inherent to the scaling properties of a time series. Indeed, the hypothesis of a fractal structure makes sense only if self-similarity is considered over several time scales across dozens of consecutive samples. A corollary of this assumption is that the process generating the time series must stay constant across the whole duration of the observation.

While the constancy assumption likely holds in treadmill walking, in normal walking, a person must regularly adjust stepping to the environment. Similarly, in cued walking experiments, it is not guaranteed that participants continuously follow cues throughout the duration of the trial. As a result, the adaptations of gait control may result in contiguous patches of strides with different fluctuation regimes. If theses patchy time series were analyzed via DFA as a whole, biased results could be obtained.

I hypothesized that false conclusions may be drawn about the fluctuation regime of a gait time series if the assumption of process constancy was violated. Therefore, the objective of this study was to explore the effects of artificial modifications of time series on DFA outcomes. I used a modeling approach to mix different proportions of correlated and anticorrelated time series. These were either purely artificial time series or actual gait data. In addition, I verified whether mixed series induced changes in correlation strengths between variability (DFA detrended variance) and time scale (DFA box size), which may help in detecting them.

## 2. Methods

The study was divided into two parts: 1) a modeling study implying artificial time series that were computer-generated; and 2) a modeling study implying real gait time series. An identical statistical treatment was used for the two parts. Matlab R2018b (Mathworks, Natick, MA) was used for all computation.

*2.1 Time series*

Artificial time series containing 10,000 samples were generated by means of autoregressive fractionally integrated moving average (ARFIMA) models. I used the script ARFIMA_SIM available on Matlab Central [15]. The auto-regressive and moving average orders of the models were set to 0 (i.e., pure long-memory signals). The differentiating parameter was set either to 0.3 or -0.2: This generated either long-range correlated time series with a scaling exponent of 0.8, or anti-correlated time series with a scaling exponent of 0.3. These values correspond to actual scaling exponents found in real gait time series (normal and cued, respectively). Eight time series were used for the subsequent analysis, i.e., four correlated series, and four anti-correlated series.

Real gait time series containing 500 consecutive strides were selected from an online dataset [16]. This set contained raw gait data collected via a treadmill instrumented with a force platform [11]. The treadmill experiment consisted of submitting individuals to either auditory cueing (metronome walking) or visual cueing (aiming with footsteps at forms projected onto the treadmill belt). Here, I exploited the auditory cueing condition and the normal (no cueing) condition. Stride time series were computed from foot pressure measures according to the method suggested by Roerdink et al [17]. I preselected four individuals with a high contrast between the two conditions. DFA of the stride durations revealed the following scaling exponents: subject #1, 0.79 and 0.28; subject #2, 0.88 and 0.25; subject #3, 0.76 and 0.23; subject #4, 0.95 and 0.33.

*2.2 In silico experiments*

The aim was to randomly mix different proportions of correlated and anti-correlated time series together (mixed time series) and then observe the resulting scaling exponents computed via DFA. DFA box sizes were set between 8 and 1/5 of series length (i.e., 2000 for artificial series and 100 for gait series). The evenly spacing algorithm was used [18]. The DFA included 32 and 15 different box sizes in artificial and gait series, respectively. An example of the analysis is shown in Fig. 1.

The mixed time series were created by randomly concatenating correlated and anti-correlated segments. The lengths of the segments were arbitrarily selected as 25, 125, and 625 for artificial time series; and as 5, 25, and 100 for the gait time series. The alternance between segments was realized as follows: 1) two time series—one correlated, and one anti-correlated—were preselected and run through according to the lengths of the segments; 2) a random number between 0 and 1 was drawn from an uniform distribution; 3) if the number was lower than a predefined proportion threshold, then an anti-correlated segment was added to the mixed time series, else, a correlated segment was added. To consider stochastic effects, 200 hybrid series were generated per condition. I explored proportion from 10% to 90% with steps of 10%.

In parallel, I collected box sizes (time scale, *n*) and detrended fluctuations [*F(n)*] for each performed DFA. As a reminder, the scaling exponent is the slope of the linear regression between *log(n)* and *log[F(n)]*. The strength of association was determined by means of Pearson's correlation coefficient. The aim was to evaluate whether mixed time series exhibited a lower correlation between *n* and *F(n)*. This was done with the understanding that it might help diagnose the presence of mixed (inconsistent) series.

*2.3 Statistics*

Overall, 43,200 mixed time series were analyzed; that is, there were 200 repetitions, 9 proportions, 3 segment lengths, 4 times series, and two sets (i.e., artificial and gait time series). Boxplots classified by proportions and segment lengths are displayed in Fig. 2 for artificial time series, and in Fig.3 for gait time series. Then, I modelized the effects of combining different proportions and segment lengths by means of multiple linear regressions. The dependent variable was the scaling exponent, and the independent variables were the proportions (treated as a continuous variable) and the segment lengths (categorical variable with 3 levels, coded as two dummy variables). A stepwise forward approach was used to select significant predictors. Full models with interactions and squared terms were retained; see supplementary material for further details. The fitted models were used to predict the most likely values and then explore the average responsiveness of the fractal index to conditions (marginal effects, Fig. 2 and 3). In addition, I predicted the effect of a modification of small proportions of segments to better illustrate the effects of sparse modifications of the time series (Tables 1 and 2).

Boxplots were used to depict the correlation results as a function of segment lengths and proportions (Fig. 4). As a reference, correlations found in pure correlated and anti-correlated series were drawn on the plots.

## 3. Results

Results evidenced a sensitivity of DFA to the mix of fluctuation regimes within the analyzed time series. The boxplots (Fig. 2 and 3) show a decrease of scaling exponent with an increasing proportion of anti-correlated segments. A large variability was observed, due to the stochastic way the mixed series were generated. The scaling exponent of a mixed time series seems not to be the weighted average of the scaling exponents of both series measured separately. For example, a 50-50 mix resulted in a scaling exponent greater than 0.65 (Fig. 2 and 3), while a value of about 0.55 ([0.8 + 0.3] / 2) would be expected.

A non-linear relationship between the proportion of anti-correlated segments and the resulting scaling-exponent of the mixed time series was noted (Fig. 2 and 3). At low proportion levels, the scaling exponent remained close to the reference value of the correlated time series. Then, with increasing proportion, the scaling exponent rapidly decreased toward the value of the anti-correlated time series. This nonlinear association was confirmed by multiple regression analysis. The stepwise procedure retained a significant quadratic term (see the online supplementary file). The effect of segment lengths used to build the mixed time series was less marked than the effect of the mixing proportions. While the segment lengths seemed to slightly shift the scaling exponent values, the decreases with proportion changes and nonlinearity were similar.

The addition of a small amount of anti-correlated segments within a correlated series had no relevant impact (Table 1). Twenty percent of inserts lowered the scaling exponents by only 3% to 5%. On the contrary, the addition of a small proportion of correlated segments within an anti-correlated series had substantial effects (Table 2). A proportion of 20% was sufficient to increase scaling exponents close to 0.5. This may result in misinterpretation of the series as random, while the series is predominantly anti-correlated.

The correlation between box sizes $n$ and detrended fluctuations $F(n)$ were not substantially impacted by the mixing of time series (Fig. 4). Regarding artificial series, the correlation coefficient fell slightly below the reference level in the high proportion of anticorrelated series, but this was likely not relevant given the small variations: The correlation coefficient stayed above 0.995 among all conditions. Regarding gait series, the correlation coefficient never fell below the reference value of the anti-correlated series.

## 4. Discussion

The assumption that inconstant time series may lead to a false conclusion about the underlying generating process is supported. Prominently, the introduction of a small proportion of correlated segments within an anti-correlated series had substantial effects. The strength of association between $n$ and $F(n)$ seemed to not be affected, and hence it was determined that it cannot be used as a means to detect a mixed series.

As shown by the boxplots and the regression results (Fig. 2 and 3, and supplementary online file), a nonlinear association was found; adding a small proportion of correlated segments into a correlated time series had stronger effects than the reverse. Twenty percent of correlated patches in an anti-correlated series was enough to reach 0.5, i.e., the value expected for an underlying random, uncorrelated process (Table 2).

In cueing experiments, if participants do not continuously synchronize their gaits, one could falsely conclude that the gait time series is the result of a random process, while it may be the result of a mix of different fluctuation regimes. The findings of some previous studies seem to support this explanation. For example, an experiment reported in a 2012 article [19] consisted of older adults walking while exposed to different modalities of auditory cueing. The DFA results showed a scaling exponent greater than 0.5 when participants walked in time with an isochronous (regular) metronome, in contrast to what would be expected. However, the participants were not specifically instructed to synchronize their gaits with the auditory stimuli, which was a condition conducive for intermittent synchronization resulting in times series of inconsistent fluctuation structures. A similar result was described in a 2016 study [20]. Note, however, that these average observations could also have been obtained by a mix of synchronizers and non-synchronizers among participants.

Anti-persistent gait patterns are associated with enhanced voluntary gait control, based on convincing evidence [11,12]. It is therefore not excluded that situations that intermittently requires a precise footstep control induce a low scaling exponent. The findings of a 2015 study support this assumption. Franz et al. [21] assessed the scaling exponents in a treadmill experiment. The participants walked while watching a speed-matched, virtual hallway with and without mediolateral visual perturbations. Under the visually perturbed condition, the scaling exponent significantly diminished in older participants; the observed average scaling exponent under visual perturbations was 0.51. One can assume that this value resulted from the mix of fractal and anti-correlated segments during the walking trial. That a constant random process was the cause of observed fluctuation patterns may not be the most likely explanation.

The results of the correlation analysis (Fig. 4) highlighted the difficulty in detecting the presence of a mixed series by examining the log-log plot of a DFA (Fig. 1). It has been described that some time series may display different scaling exponents (i.e., slopes in the log-log plot) when considering different time scales (crossover phenomena [22]). Artificial trends added to time series also induce crossovers [23]. I therefore assumed that mixing time series with different fluctuation regimes would also produce crossovers, which could be detected by a substantially lower correlation coefficient between $F(n)$ and $n$. This hypothesis was not supported.

It is probable that mixed time series also impact other methods used to assess gait complexity, such as multiscale entropy [24]. Indeed, other methods also rely on the assumption of a time-invariant underlying process that leads to long-term dependencies in gait time series. Further studies are needed to analyze this issue. Alternative methods that locally analyze signal variability may help diagnose an inconstant fluctuation regime. For example, the non-stationarity index (NSI), which estimates the dispersion of normalized local means [25], has been proven sensitive to contrasts between persistent and anti-persistent gait time series [26]. Here, too, further investigation is required to assess NSI responsiveness in the context of non-constant time series.

The fact that this study is a modeling study, and not an experiment with human participants, is both a strength and a weakness. On the strength side, artificially combining time series with different fluctuation structures excludes any experimental biases. This *in silico* approach highlights the true responsiveness of DFA to inconstant time series generated in a well-defined manner. On the weakness side, it is not expected that humans would produce intermittent time series as those generated in this study. People do not instantly change the way they walk. Adaptations to changes, for example to the appearance of an external rhythmic cue, very likely take some time. Similarly, carry-over effects are expected when the external rhythm fades. Therefore, further experimental studies are required to describe the responsiveness of DFA to actual change in gait variability patterns.

## 5. Conclusion

In studies aimed at characterizing the fluctuation structure of gait variability, the hypothesis of a continuous and steady generative process must be valid when interpreting DFA results. In cases of changes in gait control during a walking trial, the resulting time series may be a patchy ensemble of several fluctuation regimes; the resulting scaling exponent may be misinterpreted. Studies implying cueing (i.e., synchronizing footsteps with external cues) may be most at risk of suffering this issue, given that the addition of a small amount of correlated series within an anti-correlated series has a substantial effect. Further evidence from an experimental study is required to confirm these findings in human beings.

**Declarations of interest**
 None

**Acknowledgement**

This research was not supported by any specific grant from funding agencies in the public, commercial, or not-for-profit sectors.

**Author contributions**
Not applicable

**Supplementary data**
Supplementary material related to this article can be found, in the online version, at doi [to be added in the final version]

**References**

[1] V. Weerdesteyn, K.L. Hollands, M.A. Hollands, Gait adaptability, Handb Clin Neurol. 159 (2018) 135–146. https://doi.org/10.1016/B978-0-444-63916-5.00008-2.

[2] A.L. Goldberger, L.A.N. Amaral, J.M. Hausdorff, P.C. Ivanov, C.-K. Peng, H.E. Stanley, Fractal dynamics in physiology: alterations with disease and aging, Proc. Natl. Acad. Sci. U.S.A. 99 Suppl 1 (2002) 2466–2472. https://doi.org/10.1073/pnas.012579499.

[3] L.M. Decker, F. Cignetti, N. Stergiou, Complexity and Human Gait, Rev Andal Med Deporte. 3 (2010) 2–12.

[4] J.M. Hausdorff, C.K. Peng, Z. Ladin, J.Y. Wei, A.L. Goldberger, Is walking a random walk? Evidence for long-range correlations in stride interval of human gait, J. Appl. Physiol. 78 (1995) 349–358. https://doi.org/10.1152/jappl.1995.78.1.349.

[5] P. Terrier, V. Turner, Y. Schutz, GPS analysis of human locomotion: further evidence for long-range correlations in stride-to-stride fluctuations of gait parameters, Human Movement Science. 24 (2005) 97–115.

[6] B.H. Repp, Y.-H. Su, Sensorimotor synchronization: a review of recent research (2006-2012), Psychon Bull Rev. 20 (2013) 403–452. https://doi.org/10.3758/s13423-012-0371-2.

[7] M.H. Thaut, G.C. McIntosh, V. Hoemberg, Neurobiological foundations of neurologic music therapy: rhythmic entrainment and the motor system, Front Psychol. 5 (2014) 1185. https://doi.org/10.3389/fpsyg.2014.01185.

[8] S. Ghai, I. Ghai, G. Schmitz, A.O. Effenberg, Effect of rhythmic auditory cueing on parkinsonian gait: A systematic review and meta-analysis, Sci Rep. 8 (2018) 506. https://doi.org/10.1038/s41598-017-16232-5.

[9] G.E. Yoo, S.J. Kim, Rhythmic Auditory Cueing in Motor Rehabilitation for Stroke Patients: Systematic Review and Meta-Analysis, J Music Ther. 53 (2016) 149–177. https://doi.org/10.1093/jmt/thw003.

[10] P. Terrier, O. Dériaz, Persistent and anti-persistent pattern in stride-to-stride variability of treadmill walking: influence of rhythmic auditory cueing, Human Movement Science. 31 (2012) 1585–1597.

[11] P. Terrier, Fractal Fluctuations in Human Walking: Comparison Between Auditory and Visually Guided Stepping, Ann Biomed Eng. 44 (2016) 2785–2793. https://doi.org/10.1007/s10439-016-1573-y.

[12] M. Roerdink, C.P. de Jonge, L.M. Smid, A. Daffertshofer, Tightening Up the Control of Treadmill Walking: Effects of Maneuverability Range and Acoustic Pacing on Stride-to-Stride Fluctuations, Front. Physiol. 10 (2019). https://doi.org/10.3389/fphys.2019.00257.

[13] C.K. Peng, S.V. Buldyrev, A.L. Goldberger, S. Havlin, R.N. Mantegna, M. Simons, H.E. Stanley, Statistical properties of DNA sequences, Physica A. 221 (1995) 180–192.

[14] V. Marmelat, R.L. Meidinger, Fractal analysis of gait in people with Parkinson's disease: three minutes is not enough, Gait Posture. 70 (2019) 229–234. https://doi.org/10.1016/j.gaitpost.2019.02.023.

[15] S. Fatichi, ARFIMA simulations, 2009. https://ch.mathworks.com/matlabcentral/fileexchange/25611-arfima-simulations.

[16] P. Terrier, Complexity of human walking: the attractor complexity index is sensitive to gait synchronization with visual and auditory cues, (2019). https://doi.org/10.6084/m9.figshare.8166902.v2.

[17] M. Roerdink, B.H. Coolen, B.H.E. Clairbois, C.J.C. Lamoth, P.J. Beek, Online gait event detection using a large force platform embedded in a treadmill, J Biomech. 41 (2008) 2628–2632. https://doi.org/10.1016/j.jbiomech.2008.06.023.

[18] Z.M.H. Almurad, D. Delignières, Evenly spacing in Detrended Fluctuation Analysis, Physica A: Statistical Mechanics and Its Applications. 451 (2016) 63–69. https://doi.org/10.1016/j.physa.2015.12.155.

[19] J.P. Kaipust, D. McGrath, M. Mukherjee, N. Stergiou, Gait variability is altered in older adults when listening to auditory stimuli with differing temporal structures, Ann Biomed Eng. 41 (2013) 1595–1603. https://doi.org/10.1007/s10439-012-0654-9.

[20] N.K. Bohnsack-McLagan, J.P. Cusumano, J.B. Dingwell, Adaptability of stride-to-stride control of stepping movements in human walking, Journal of Biomechanics. 49 (2016) 229–237. https://doi.org/10.1016/j.jbiomech.2015.12.010.

[21] J.R. Franz, C.A. Francis, M.S. Allen, S.M. O'Connor, D.G. Thelen, Advanced age brings a greater reliance on visual feedback to maintain balance during walking, Human Movement Science. 40 (2015) 381–392. https://doi.org/10.1016/j.humov.2015.01.012.

[22] A. Muñoz-Diosdado, L. Guzmán-Vargas, A. Ramírez-Rojas, J.L. Del Río-Correa, F. Angulo-Brown, Some cases of crossover behavior in heart interbeat and electroseismic time series, Fractals. 13 (2005) 253–263. https://doi.org/10.1142/S0218348X05002970.

[23] K. Hu, P.Ch. Ivanov, Z. Chen, P. Carpena, H. Eugene Stanley, Effect of trends on detrended fluctuation analysis, Phys. Rev. E. 64 (2001) 011114. https://doi.org/10.1103/PhysRevE.64.011114.

[24] M. Costa, C.-K. Peng, A. L. Goldberger, J.M. Hausdorff, Multiscale entropy analysis of human gait dynamics, Physica A: Statistical Mechanics and Its Applications. 330 (2003) 53–60. https://doi.org/10.1016/j.physa.2003.08.022.

[25] J.M. Hausdorff, A. Lertratanakul, M.E. Cudkowicz, A.L. Peterson, D. Kaliton, A.L. Goldberger, Dynamic markers of altered gait rhythm in amyotrophic lateral sclerosis, J. Appl. Physiol. 88 (2000) 2045–2053. https://doi.org/10.1152/jappl.2000.88.6.2045.

[26] P. Terrier, Step-to-step variability in treadmill walking: influence of rhythmic auditory cueing, PLoS ONE. 7 (2012) e47171. https://doi.org/10.1371/journal.pone.0047171.

# Tables

**Table 1**

Effect of insertion of a small proportion of anti-correlated time series within a correlated time series.

| Segment lengths | Proportion of anti-correlated segments within correlated time series | | | | | | | |
|---|---|---|---|---|---|---|---|---|
| | 5% | | 10% | | 15% | | 20% | |
| | absolute | relative | absolute | relative | absolute | relative | absolute | relative |
| 5 | 0.00 | -0.6% | -0.01 | -1.5% | -0.02 | -2.6% | -0.03 | -4.0% |
| 25 | 0.00 | -0.4% | -0.01 | -1.0% | -0.02 | -1.8% | -0.02 | -2.9% |
| 100 | -0.01 | -0.9% | -0.02 | -2.0% | -0.03 | -2.4% | -0.04 | -5.0% |

A multiple regression model (see supplementary files) was used to predict absolute and relative (percent change) differences among time series without insertion and time series including 5% to 20% anti-correlated segments. The regression model was based on actual gait data.

**Table 2**

Effect of insertion of a small proportion of correlated time series within an anti-correlated time series.

| Segment lengths | Proportion of correlated segments within anti-correlated time series | | | | | | | |
|---|---|---|---|---|---|---|---|---|
| | 5% | | 10% | | 15% | | 20% | |
| | absolute | relative | absolute | relative | absolute | relative | absolute | relative |
| 5 | 0.04 | 13.8% | 0.08 | 27.0% | 0.12 | 39.5% | 0.16 | 51.4% |
| 25 | 0.04 | 10.4% | 0.08 | 20.2% | 0.12 | 29.6% | 0.15 | 38.4% |
| 100 | 0.05 | 15.1% | 0.09 | 29.6% | 0.13 | 43.3% | 0.17 | 56.4% |

A multiple regression model (see supplementary files) was used to predict absolute and relative (percent change) differences among time series without insertion and time series including 5% to 20% correlated segments. The regression model was based on actual gait data.

# Figures

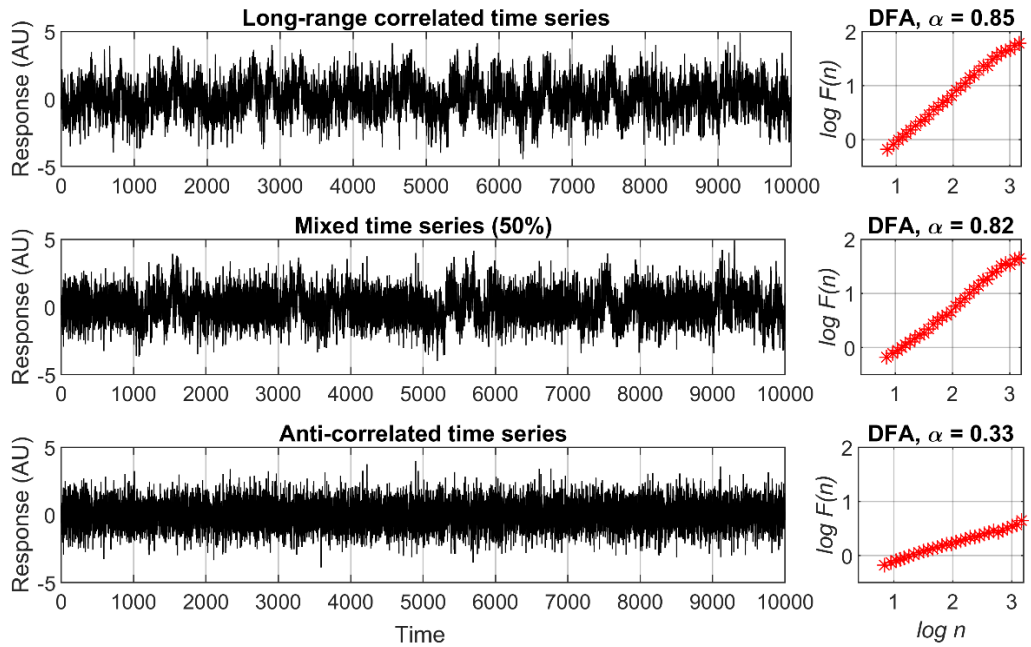

**Fig. 1.** Illustration of the method for analysing the mixed time series. Left: Three different types of artificial time series (autoregressive fractionally integrated moving average [ARFIMA] signals); Top: Pure long-range correlated (fractal) time series; Bottom: Pure anti-correlated time-series; Middle: Mix of the top and bottom series with a 50% proportion; the response is given in arbitrary units (AU). Right: Detrended fluctuation analysis (DFA) of the adjacent time series. Detrended fluctuation [F(n)] and box size (time scale, n) are plotted in a log-log plot. The scaling exponent α is the slope of the linear fit.

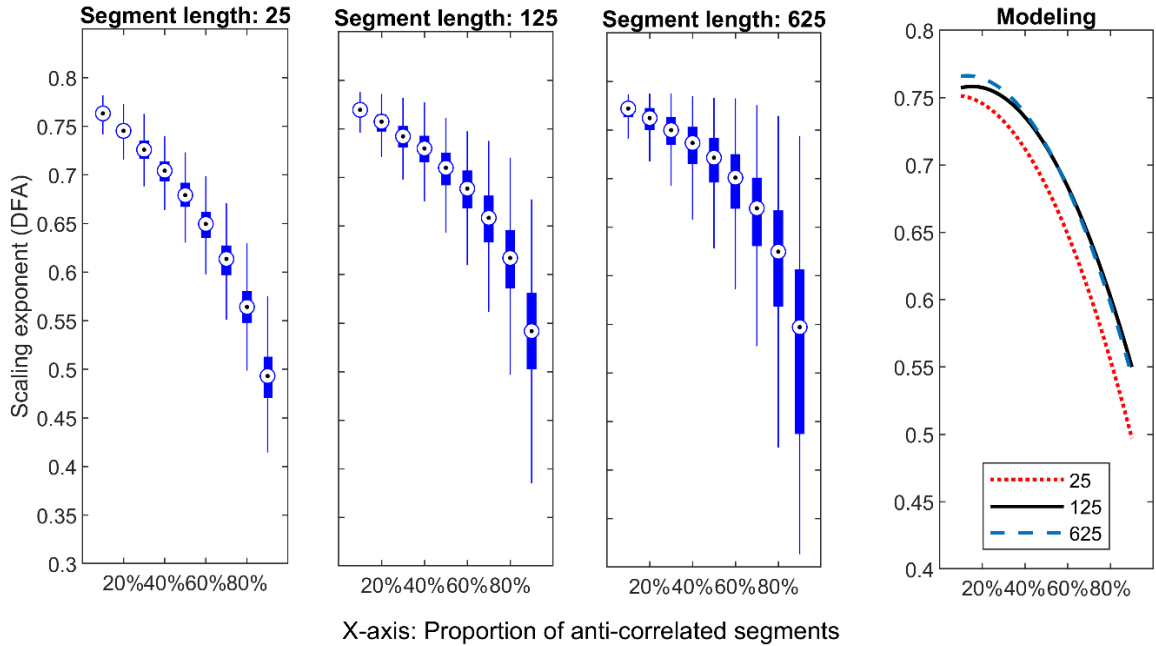

**Fig. 2.** Artificial mixed time series. Different proportions (10% to 90%) of anti-correlated segments of 25, 125, and 625 samples were randomly inserted within a long-range correlated time series of 10,000 samples. Each box plot represents the median (circle), quartiles (box), and data extent (lines) of 200 realizations of the random mixing process. The plot on the right shows the results of the multiple regression model (marginal effects).

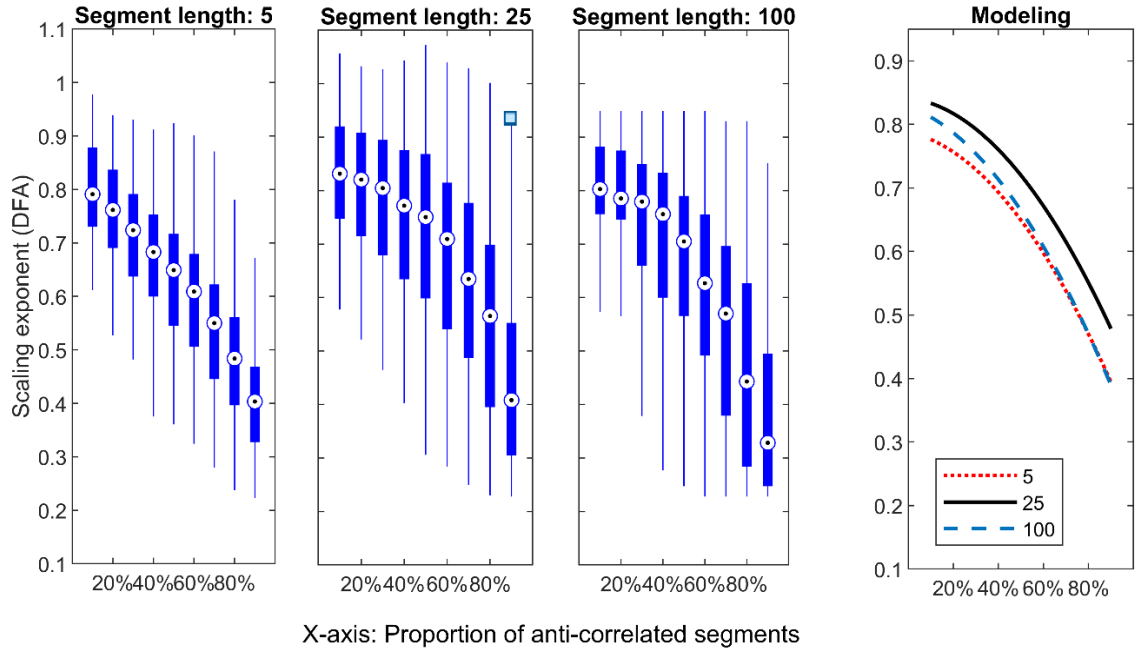

**Fig. 3.** Gait mixed time series. Different proportions (10% to 90%) of anti-correlated segments of 5, 25, and 100 samples were randomly inserted within a long-range correlated time series of 500 samples. Each box plot represents the median (circle), quartiles (box), and data extent (lines) of 200 realizations of the random mixing process. The plot on the right shows the result of the multiple regression model (marginal effects).

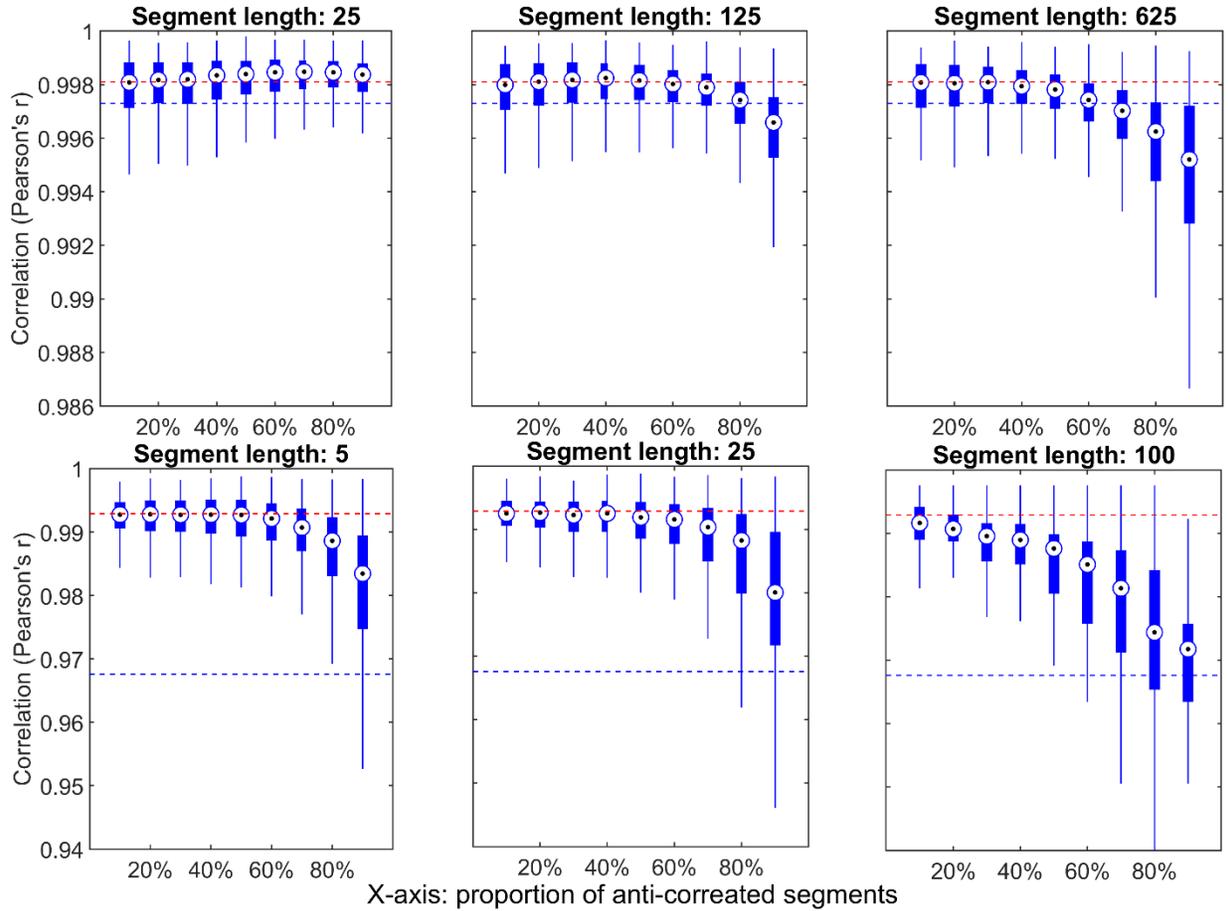

**Fig 4.** Correlation analysis. Correlation coefficients (Pearson's r) between *F(n)* and *n* (see Fig. 1) were computed for each DFA, as presented in Fig. 2 and 3. The top row shows the results for the artificial mixed time series. The bottom row shows the results for the gait mixed time series. Each box plot represents the median (circle), quartiles (box), and data extent (lines) of 200 realizations of the random mixing process. Dashed lines show the r value for original (without mixing) correlated time series (red) and original anti-correlated time series (blue).